\documentclass[11pt]{article}
\pdfoutput=1
\baselineskip
\usepackage{amssymb, graphicx, latexsym, cite, collref, epsfig, amsmath, bbm, color, hyperref, rotating}
\newcommand{\Z}{\mathbb{Z}}
\newcommand{\R}{\mathbb{R}}

\newcommand{\SU}{\mathrm{SU}}
\newcommand{\U}{\mathrm{U}}
\newcommand{\SO}{\mathrm{SO}}

\newcommand{\ud}{\mathop{}\!\mathrm{d}}

\newcommand{\SNG}{S_{\mbox{\tiny{NG}}}}
\newcommand{\Scl}{S_{\mbox{\tiny{cl}}}}
\newcommand{\SPol}{S_{\mbox{\tiny{Pol}}}}
\newcommand{\Sb}{S_{\mbox{\tiny{b}}}}
\newcommand{\VNG}{V_{\mbox{\tiny{NG}}}}
\newcommand{\Vb}{V_{\mbox{\tiny{b}}}}
\newcommand{\Vext}{V_{\mbox{\tiny{ext}}}}
\newcommand{\QNG}{Q_{\mbox{\tiny{NG}}}}
\newcommand{\Qb}{Q_{\mbox{\tiny{b}}}}
\newcommand{\Qr}{Q_{\mbox{\tiny{r}}}}
\newcommand{\Rmin}{R_{\mbox{\tiny{min}}}}
\newcommand{\gammaE}{\gamma_{\mbox{\tiny{E}}}}
\newcommand{\re}{{\rm{Re}}}

\newcommand{\tr}{{\rm Tr\,}}

\newcommand{\redchisq}{\chi^2_{\tiny\mbox{red}}}
\newcommand{\eq}{\begin{equation}}
\newcommand{\en}{\end{equation}}
\newcommand{\eqar}{\begin{eqnarray}}
\newcommand{\enar}{\end{eqnarray}}

\begin{document}
\begin{titlepage}
\begin{flushright} 
IFT-UAM/CSIC-14-052
\end{flushright} 
\vskip0.5cm 
\begin{center}
{\Large\bf A different kind of string}
\end{center}
\vskip1.0cm
\centerline{Michele~Caselle$^{a}$, Marco~Panero$^{a,b}$, Roberto~Pellegrini$^{c}$ and Davide Vadacchino$^{a}$}
\vskip0.5cm
\centerline{\sl $^a$ Dipartimento di Fisica dell'Universit\`a di Torino and INFN, Sezione di Torino,}
\centerline{\sl Via P.~Giuria 1, I-10125 Torino, Italy}
\centerline{\sl $^b$ Instituto de F\'{\i}sica T\'eorica UAM/CSIC, Universidad Aut\'onoma de Madrid}
\centerline{\sl Calle Nicol\'as Cabrera 13-15, Cantoblanco E-28049 Madrid, Spain}
\centerline{\sl $^c$ Physics Department, Swansea University, Singleton Park, Swansea SA2 8PP, UK}
\begin{center}
{\sl  E-mail: }  
\href{mailto:caselle@to.infn.it}{\texttt{caselle@to.infn.it}},  \href{mailto:panero@to.infn.it}{\texttt{panero@to.infn.it}}, \href{mailto:ropelleg@to.infn.it}{\texttt{ropelleg@to.infn.it}}, \href{mailto:vadacchi@to.infn.it}{\texttt{vadacchi@to.infn.it}}
\end{center}
\vskip1.0cm
\begin{abstract}
In $\U(1)$ lattice gauge theory in three spacetime dimensions, the problem of confinement can be studied analytically in a semi-classical approach, in terms of a gas of monopoles with Coulomb-like interactions. In addition, this theory can be mapped to a spin model via an exact duality transformation, which allows one to perform high-precision numerical studies of the confining potential. Taking advantage of these properties, we carried out an accurate investigation of the effective string describing the low-energy properties of flux tubes in this confining gauge theory. We found striking deviations from the expected Nambu-Goto-like behavior, and, for the first time, evidence for contributions that can be described by a term proportional to the extrinsic curvature of the effective string worldsheet. Such term is allowed by Lorentz invariance, and its presence in the infrared regime of the $\U(1)$ model was indeed predicted by Polyakov several years ago. Our results show that this term scales as expected according to Polyakov's solution, and becomes the dominant contribution to the effective string action in the continuum limit. We also demonstrate analytically that the corrections to the confining potential induced by the extrinsic curvature term can be related to the partition function of the massive perturbation of a $c=1$ bosonic conformal field theory. The implications of our results for $\SU(N)$ Yang-Mills theories in three and in four spacetime dimensions are discussed.
\end{abstract}
\vspace*{0.2cm}

\end{titlepage}

\section{Introduction}
\label{sect:intro}

One of the most interesting recent results in the effective string description of the dynamics of long flux tubes in confining Yang-Mills theories is the proof of universality of the first few terms in their effective action. This is a direct consequence of the symmetry constraints one must impose on the action, and makes this effective theory much more predictive than other effective models in particle physics. These constraints were first obtained by comparing the string partition function in different channels (``open-closed string duality'')~\cite{Luscher:2004ib, Aharony:2009gg}. However, it was later realized~\cite{Meyer:2006qx, Aharony:2011gb, Gliozzi:2011hj, Gliozzi:2012cx, Meineri:2013ew} that there was a simpler way to understand these constraints and that they are, in fact, a direct consequence of Poincar\'e symmetry in the underlying Yang-Mills theory. There are two main routes one can follow to impose this symmetry in the effective action. The first is to keep the original string action, without fixing the reparametrization invariance. This approach is not the simplest one to perform calculations, but it allows a better understanding of the various terms which appear in an expansion around the long-string limit. In this framework the effective action is obtained by the mapping 
\begin{equation}
X^\mu : \mathcal{M} \rightarrow \R^D, \qquad \mu = 0, \cdots , D-1
\end{equation}
of the two-dimensional surface describing the worldsheet of the string $\cal M$ into 
the (flat) $D$-dimensional target space $\R^D$ of the gauge theory (here and in the following, we assume Euclidean signature for both the worldsheet and the target space) and then imposing the constraint due to Poincar\'e and parity invariance of the original theory. This approach was discussed in detail in ref.~\cite{Aharony:2013ipa}. The first few terms of the action compatible with these constraints are suitable combinations of geometric invariants, which can be  constructed from the induced metric $g_{\alpha\beta}=\partial_\alpha X^\mu\partial_\beta X_\mu$. These terms can be classified according to their ``weight'', defined as the difference between the number of derivatives minus the number of fields $X^\mu$ (i.e., as their energy dimension). Due to invariance under parity, only terms with an even number of fields should be considered. The only term of weight zero corresponds to the well-known Nambu-Goto (NG) action
 \begin{equation}
 \SNG=\sigma\int d^2\xi \sqrt{g}~,
 \end{equation}
where $g \equiv \det(g_{\alpha\beta})$ and we have denoted the worldsheet coordinates as $\xi\equiv(\xi^0,\xi^1)$. This term has a natural geometric interpretation: it measures the area of the string worldsheet. At weight two, two new contributions appear:
\begin{eqnarray}
&& S_{2,\cal{R}}=\gamma\int d^2\xi \sqrt{g} \cal{R}, \\
&& S_{2,K}=\alpha\int d^2\xi \sqrt{g} K^2, \label{ext}
\end{eqnarray}
where $\sigma$, $\alpha$, and $\gamma$ are the only free parameters of the effective theory up to this level, $\cal{R}$ denotes the Ricci scalar constructed from the induced metric, and $K$ is the extrinsic curvature, defined as $K=\Delta (g) X$, with  
\begin{equation}
\Delta(g)=\frac{1}{\sqrt{(g)}}\partial_a[\sqrt{(g)}g^{ab}\partial_b]
\end{equation}
the Laplacian in the space with metric $g_{\alpha\beta}$.

At weight four, several new combinations can be constructed, including for instance a term proportional to the square of the Ricci scalar. The argument which is used at this point to further  constrain the effective action is that the term proportional to $\cal{R}$ is topological in two dimensions and, since in the long-string limit we are interested in one does not expect topology-changing fluctuations, its contribution can be neglected~\cite{Aharony:2013ipa}. On the other hand, the term in eq.~(\ref{ext}) which contains $K^2$ is proportional to the equation of motion of the Nambu-Goto Lagrangian and can be eliminated by a suitable field redefinition. Hence also this term can be neglected---at least from a classical point of view.

The same result can be obtained following the second of the two routes mentioned above, fixing the reparametrization invariance to the unitary gauge (often called ``physical gauge''), which we are going to assume in the following. In this gauge the two worldsheet coordinates are identified with the longitudinal degrees of freedom (d.o.f.) of the string: $\xi^0=X^0$, $\xi^1=X^1$, so that the string action can be expressed as a function of the $(D-2)$ d.o.f. corresponding to the transverse displacements of the string worldsheet, $X^i$, with $i=2, \dots , (D-1)$. With this choice, one is neglecting worldsheet configurations corresponding to ``back-tracking'' or self-intersecting surfaces. This is a good approximation for applications in the infrared dynamics of confining gauge theories, and it makes the transverse displacements single-valued functions of the worldsheet coordinates (thereby simplifying analytical computations). This restriction can be interpreted as an analogue of the aforementioned assumption, that the Ricci scalar is constant for the string worldsheet surfaces.

The action can then be written as a low-energy expansion in the number of derivatives of the transverse d.o.f. of the string. The first few terms in this expansion are
\begin{equation}
S=\Scl+\frac\sigma2\int d^2\xi\left[\partial_\alpha X_i\cdot\partial^\alpha X^i+
c_2(\partial_\alpha X_i \cdot\partial^\alpha X^i)^2
+c_3(\partial_\alpha X_i \cdot\partial_\beta X^i)^2+\dots\right],
\label{action}
\end{equation}
where the classical action $\Scl$ includes the terms corresponding to the minimal area of the string worldsheet (and possibly a perimeter term), while the second term describes a massless free field theory in two dimensions~\cite{Luscher:1980fr} and subsequent terms correspond to higher-order interactions among the $X_i$ fields. 

The main point in this derivation is that the $c_i$ coefficients are not completely arbitrary, but must satisfy a set of constraints to enforce Lorentz invariance of the theory. In fact, even though the $\SO(D)$ invariance of the original theory is spontaneously broken by the formation of the classical string configuration around which one is expanding, the effective action should still respect this symmetry through a non-linear realization in terms of the transverse fields $X_i$~\cite{Meyer:2006qx, Aharony:2011gb, Gliozzi:2011hj, Gliozzi:2012cx, Meineri:2013ew}. These non-linear constraints induce a set of recursive relations among the coefficients of the expansion, which strongly reduce the number of free parameters of the theory. In particular, it can be shown that the terms with only first derivatives coincide with the Nambu-Goto action to all orders in the derivative expansion~\cite{Aharony:2010cx}. Imposing these constraints to the next-to-leading-order terms (i.e. to the terms beyond the Nambu-Goto action) in the effective string action, one obtains the gauge-fixed version of the two contributions $S_{2,\mathcal{R}}$ and $S_{2,K}$ mentioned above~\cite{Gliozzi:2012cx, Meineri:2013ew}. At this point, using the Nambu-Goto equations of motion and a suitable redefinition of the $X_i$ fields, it is possible to show that, for the ground-state quark-antiquark potential $V(R)$ (where $R$ denotes the distance between the static color sources, which is taken to be large) in a confining theory in three spacetime dimensions, the first deviation with respect to the prediction given by the pure Nambu-Goto action appears only at $O(R^{-7})$.
  
The fact that the first deviations from the Nambu-Goto string appear at such a high order explains why earlier Monte Carlo calculations~\cite{Caselle:1994df, Caselle:2005xy} found good agreement with the predictions of a Nambu-Goto string.

However, thanks to the improvement in the accuracy of simulations, during the past few years it has become possible to observe deviations from the expected ``universal'' behavior~\cite{Athenodorou:2010cs, Caselle:2007yc, Caselle:2010pf, Brandt:2010bw}. These deviations were observed both in the excited string states of $\SU(N)$ Yang-Mills theories~\cite{Athenodorou:2010cs, Brandt:2010bw} and in the ground-state potential in the $\Z_2$ lattice gauge theory in three dimensions (3D). In the latter case, deviations were observed for worldsheets with torus~\cite{Caselle:2007yc} and cylinder topology~\cite{Caselle:2010pf}. For recent reviews of lattice studies about these subjects, see ref.~\cite[sub-subsections (5.1.3) and (5.2.2)]{Lucini:2012gg} and ref.~\cite{Brandt:2013eua}.

In all of the cases above, these corrections were rather small and required high-precision lattice simulations to be observed. By contrast, the situation in the 3D $\U(1)$ theory is dramatically different: as it will be shown in detail in the present article, macroscopic deviations from the expectations described above can be observed for a wide range of distances and values of the Wilson action parameter $\beta=1/(a e^2)$ (where $a$ denotes the lattice spacing and $e$ is the coupling, which in 3D has energy dimension $1/2$)~\cite{Vadacchino:2013baa}. These deviations turn out to be incompatible with the expected, $O(R^{-7})$ terms. These results prompted us to reconsider the various steps of the above analysis, and led us to realize that the field redefinition needed to eliminate the extrinsic curvature is anomalous and that this term, which can be indeed eliminated at tree level, may give a non trivial contribution at one loop that, in a certain range of values of $R$, could be more important than the $O(R^{-7})$ correction mentioned above.

In this article we present a complete set of novel numerical results for the confining potential in the 3D $\U(1)$ theory, and show that they can be described for all $\beta$ values by including the extrinsic-curvature contribution in the effective string action. This confirms for the first time an earlier theoretical prediction by Polyakov, who first suggested the presence of an extrinsic-curvature term in the effective string action for this model in ref.~\cite{Polyakov:1996nc}. Our lattice results confirm his prediction and allow to quantify its effect in the interquark potential.

This article is organized as follows. In the next section the main properties of the $\U(1)$ model in three spacetime dimensions are discussed. In sect.~\ref{sect:string} we review the basics of the effective string description for confining gauge theories and discuss the zeta-function regularization of the extrinsic-curvature term. Then, in sect.~\ref{sect:results} we present our new Monte Carlo results and compare them with the effective string prediction. The last section~\ref{sect:conclusions} includes comments on the implications of our results for non-Abelian $\SU(N)$ gauge theories and some concluding remarks. Finally, in the appendix~\ref{app:reabsorbing} we show that, although classically the extrinsic-curvature term in the action can be reabsorbed into the Gaussian one through a simple field redefinition, the latter reproduces the rigid-string contribution via quantum effects at one loop.

\section{$\U(1)$ gauge theory in three spacetime dimensions}
\label{sect:U1}

In this section we summarize some well-know facts about the $\U(1)$ gauge theory in three spacetime dimensions, and the lattice regularization thereof, defined by the Wilson action~\cite{Wilson:1974sk}
\begin{equation}
\label{Wilson_action}
\beta \sum_{x \in \Lambda} \sum_{1 \le \mu < \nu \le 3} \left[1 - \re \, U_{\mu\nu}(x)\right], \;\;\; \mbox{with} \;\;\; \beta=\frac{1}{a e^2},
\end{equation}
where $\Lambda$ denotes an isotropic cubic lattice of spacing $a$, $e$ is the bare lattice coupling, and 
\begin{equation}
\label{plaquette}
U_{\mu\nu}(x) = U_\mu(x) U_\nu(x+a\hat\mu) U^\star_\mu(x+a\hat\nu) U^\star_\nu(x), \;\;\; \mbox{with} \;\;\; U_\mu(x) = \exp \left[ i a A_\mu \left( x + a \hat{\mu}/2 \right) \right].
\end{equation}
The remarkable feature of this theory is that it can be studied analytically in the semi-classical approximation~\cite{Polyakov:1976fu, Gopfert:1981er}: one can show that the model is confining for all values of $\beta$, and that in the $ \beta\gg 1$ limit it reduces to a theory of free massive scalars. In this limit, the mass of the lightest glueball and the string tension (in lattice units) behave as
\begin{equation}
m_0 a = c_0\sqrt{8 \pi^2 \beta}e^{-\pi^2 v(0)\beta},\;\;\; \sigma a^2 \geq
\frac{c_{\sigma}}{\sqrt{2\pi^2\beta}} e^{-\pi^2 v(0)\beta},
\end{equation}
where $v(0) \simeq 0.2527 \dots $ denotes the zero-distance Coulomb potential in lattice units, and in the semi-classical approximation $c_0=1$ and $c_{\sigma}=8$. Previous numerical studies~\cite{Loan:2002ej} (which our simulations confirm) showed that the string tension saturates this bound and that both constants are affected by the semi-classical approximation, changing their values in the continuum limit. Despite these quantitative differences, both $m_0$ and $\sigma$ remain strictly positive, so the model is confining at any value of $\beta$. The point in using such lattice model at finite spacing is that, while in general for confining lattice gauge theories the $m_0/\sqrt{\sigma}$ ratio is approximately fixed (up to discretization effects), in this model we have
\begin{equation}
\label{ratio_m0}
\frac{m_0}{\sqrt{\sigma}} = \frac{2 c_0}{\sqrt{c_\sigma}} 
(2\pi^2\beta)^{3/4} e^{-\pi^2 v(0)\beta/2},
\end{equation}
so, by changing $\beta$, we can tune the $m_0/\sqrt{\sigma}$ ratio of the lattice theory to any chosen value.

\subsection{Duality transformation}
\label{subsect:duality}

Since the model is invariant under an Abelian gauge symmetry, one can easily perform a duality transformation~\cite{Kramers:1941kn} (see also refs.~\cite{Wegner:1984qt, Wegner:2014ixa} for a discussion) and obtain a simple spin model with global $\mathbb{Z}$ symmetry and integer-valued ${}^\star s$ variables, defined on the sites of the dual lattice (note that, in $D=3+1$ dimensions, the same transformation leads to a model with local $\mathbb{Z}$ symmetry~\cite{Zach:1997yz, Panero:2005iu, Panero:2004zq, Cobanera:2011wn, Mercado:2013ola}). More precisely, the duality transformation is an \emph{exact} map of the original partition function to
\begin{equation}
Z = \sum_{ \{ {}^\star s \in \Z \} } \prod_{\mbox{\tiny{links}}} I_{|d {}^\star s|} (\beta),
\end{equation}
where $I_\nu(z)$ denotes the modified Bessel function of the first kind of order $\nu$, the product runs over the elementary links of the dual lattice, and $d {}^\star s$ denotes the difference between ${}^\star s$ variables at the ends of a link. 

The dual formulation of the system has several advantages over the original one. First of all, from the computational point of view, the model is much easier and faster to simulate, since we deal with a spin model. Moreover, a $Q\bar{Q}$ pair of static sources (at a distance $R$ from each other) can be easily included in the partition function of the dual model, which then takes the form
\begin{equation}
Z_R  = \sum_{ \{ {}^\star s \in \Z \} } \prod_{\mbox{\tiny{links}}} I_{|d {}^\star s+ {}^\star n|} (\beta),
\end{equation}
where ${}^\star n$ is an integer-valued 1-form which must be non-vanishing on a set of links dual to an arbitrary surface bounded by the two loops (in our implementation, we chose the surface of minimal area).

As a consequence, the two-point correlation function of Polyakov loops $P$ can be written as
\begin{equation}
\label{Pol_correlator_in_dual}
\langle P^\star (R) P(0) \rangle = \frac{Z_R}{Z}.
\end{equation}
For large $R$, the quantity appearing on the right-hand side of eq.~(\ref{Pol_correlator_in_dual}) is the ratio of two partition functions dominated by poorly overlapping sets of typical configurations, and, as a consequence, computing the expectation value of the $Q\bar{Q}$ pair involves the numerical challenge of an exponentially decaying signal-to-noise ratio. This happens both in the original and in the dual formulation. However, in the dual formulation this problem can be bypassed, as the ratio on the right-hand side of eq.~(\ref{Pol_correlator_in_dual}) can be factorized into a product of ratios, in which each numerator is the partition function of a system differing from the one described by the denominator by the insertion of a non-vanishing ${}^\star n$ only on one link. Denoting the length of the Polyakov loops and their separation in units of the lattice spacing as $n_t$ and $n_R$ respectively, one obtains:
\begin{equation}
\label{snake}
\langle P^\star (R) P(0) \rangle = \prod_{i=0}^{n_R n_t -1} \frac{Z^{(i+1)}}{Z^{(i)}}.
\end{equation}
This factorization, first proposed in the computation of 't~Hooft loops in $\SU(2)$ Yang-Mills theory in ref.~\cite{deForcrand:2000fi}, goes under the name of ``snake algorithm'', and allows one to reconstruct $\langle P^\star (R) P(0) \rangle$ from the product of factors which are not affected by a severe overlap problem (since the partition functions of systems which differ by the insertion of only one additional non-vanishing ${}^\star n$ are dominated by contributions from largely overlapping sets of typical configurations).

In addition, the efficiency in the numerical computation of the $Z^{(i+1)}/Z^{(i)}$ ratios on the right-hand side of eq.~(\ref{snake}) can be easily improved by means of a hierarchical update scheme (in which, taking advantage of the locality of the theory, portions of the lattice in the neighborhood of the ${}^\star n$ term, by which numerator and denominator differ, are updated more often), as was done for the 3D $\Z_2$ gauge theory in ref.~\cite{Caselle:2002ah}.

\subsection{A rigid-string description for the $\U(1)$ model}

Besides its expediency for numerical computations, the other major advantage of the duality transformation is that it gives insight into the physical mechanism driving confinement. Indeed, it reveals that confinement in the 3D $\U(1)$ gauge model is due to the condensation of monopole configurations~\cite{Polyakov:1976fu}. The remarkable success of this approach led to conjecture that a similar mechanism could drive confinement also in non-Abelian Yang-Mills theories, including, in particular, in the $\SU(3)$ theory in four spacetime dimensions. According to this conjecture (known as the ``dual superconductor picture''), quarks are confined by vortex lines which behave as strings.

The implicit assumption behind this scenario is that there should exist a duality transformation mapping gauge fields into strings. In the non-Abelian case, such gauge/string duality transformation is in general unknown,\footnote{A notable exception, however, is given by the holographic correspondence, relating gauge theories and string theories defined in a higher-dimensional spacetime~\cite{Maldacena:1997re, Gubser:1998bc, Witten:1998qj}.} but in the 3D $\U(1)$ case Polyakov~\cite{Polyakov:1996nc} (see also~\cite{Antonov:1998kw} for an alternative derivation) was able to give a heuristic proof of this mapping and proposed to describe the free energy of a large Wilson loop with a string action combining both the Nambu-Goto and the extrinsic curvature terms (the so called ``rigid string''). Polyakov was also able to compute the dependence of their coupling constants on the electric charge and the glueball mass of the original $\U(1)$ theory (see ref.~\cite[eq.~(17)]{Polyakov:1996nc}):
\begin{equation}
\SPol= c_1 e^2 m_0 \int d^2\xi \sqrt{g}~ + c_2 \frac{e^2}{m_0} \int d^2\xi \sqrt{g} K^2,
\label{Polyakov}
\end{equation}
where $c_1$ and $c_2$ are two undetermined constants. If we identify these coupling constants with $\sigma$ and $\alpha$ defined above, we find (apart from an undetermined constant) $\sqrt{\sigma/\alpha} \sim m_0$. This result will play an important r\^ole in the following.

\section{Effective string action and extrinsic curvature}
\label{sect:string}

Following the discussion of sect.~\ref{sect:intro}, the most general effective string action involving only terms up to weight 2 respecting Poincar\'e invariance is
\begin{equation}
S = \SNG + S_{2,K} + \Sb,
\label{action2}
\end{equation}
where, as explained in sect.~\ref{sect:intro}, we neglected the term proportional to the Ricci scalar, but included the term proportional to the square of the extrinsic curvature. 

In addition, we also included a boundary term $\Sb$, which describes the interaction of the effective string with the Polyakov loops. Also the form of $\Sb$ is strongly constrained by  Poincar\'e invariance: if the boundary is a Polyakov line in the $\xi_0$ direction located at $\xi_1=0$, for which we assume  Dirichlet boundary conditions, $X_i(\xi_0,0)=0$, then, in the physical gauge $\Sb$ can be expanded as
\begin{equation}
\Sb=\int d\xi_0 \left[b_1\partial_1 X_i\cdot \partial_1 X^i+b_2\partial_1\partial_0 X_i\cdot \partial_1\partial_0 X^i+ \dots\right].
\label{bounda}
\end{equation}
Imposing Lorentz invariance one can show that $b_1=0$~\cite{Luscher:2004ib, Aharony:2010cx} and that the term having $b_2$ as its coefficient is nothing but the first contribution arising from the Lorentz-invariant combination~\cite{Billo:2012da}
\begin{equation}
b_2\int d\xi_0 \left[\frac{\partial_0\partial_1 X_i\cdot\partial_0\partial_1 X^i}{1+\partial_1 X_i\cdot\partial_1X^i}-\frac{\left(\partial_0\partial_1 X_i\cdot\partial_1 X^i\right)^2}{\left(1+\partial_1 X_i\cdot\partial_1X^i\right)^2}\right].
\label{firstb}
\end{equation} 
The precision of our data is sufficient to identify finite-size corrections in the interquark potential up to $O(R^{-4})$: hence, we truncate the expansion of eq.~(\ref{action2}) in powers of $X$ to the corresponding order. Moreover, from now on we restrict our attention to the $D=3$ case, so that the transverse displacement of the string from its classical configuration  is described by a single bosonic field $X(\xi_1,\xi_2)$. We find:\footnote{In eq.~(\ref{action4}) we truncate the rigidity term to the Gaussian part, since higher-order terms give corrections beyond our resolution.}
\begin{eqnarray}
&& \SNG \simeq 
\Scl+\frac\sigma2\int d^2\xi\left[\partial_\alpha X\cdot\partial^\alpha X-
\frac14(\partial_\alpha X \cdot\partial^\alpha X)^2\right],
\label{action3} \\
&& S_{2,K} \simeq \alpha\int d^2\xi (\Delta X)^2,
\label{action4} \\
&& \Sb \simeq b_2\int d\xi_0 \left[\partial_1\partial_0 X\cdot \partial_1\partial_0 X
\right]. \label{bounda2}
\end{eqnarray}
Thus we are left with three free parameters ($\sigma$, $\alpha$ and $b_2$) which will be fitted comparing with the numerical data.

The action in eq.~(\ref{action2}) has a long history. Originally introduced to describe the physics of fluid membranes~\cite{Hel85, Pel85, Forster1986115}, it was later proposed by Polyakov and by Kleinert as a way to stabilize the Nambu-Goto action~\cite{Polyakov:1986cs, Kleinert:1986bk}. Its contribution to the interquark potential was evaluated in the large-$D$ limit in ref.~\cite{Braaten:1986bz}, and then for generic $D$ in ref.~\cite{German:1989vk}. The corrections induced in higher-order terms of the spectrum have been recently evaluated in ref.~\cite{Ambjorn:2014rwa}. 

In the next subsection, we will re-discuss this action from a slightly different point of view.

\subsection{Zeta-function regularization of the extrinsic curvature action}

The contribution of the extrinsic-curvature term to the interquark potential can be evaluated using the zeta-function regularization~\cite{Nesterenko:1997ku}. Let us review the main steps of this calculation. Let us first concentrate on the contribution due to the Gaussian integration over transverse d.o.f.: the Gaussian part of the action is
\begin{equation}
\label{Gaussian_part_of_action}
S=\sigma\int\limits_{0}^{N_t}dt\int\limits_{0}^{R}dr\left[1+\frac{1}{2}\partial_\alpha X\cdot\partial^\alpha X\right] +
\alpha\int\limits_{0}^{N_t}dt\int\limits_{0}^{R}dr~(\Delta X)^2,
\end{equation}
where $R$ denotes the interquark distance, $N_t$ is the system size in the Euclidean time direction (i.e. the length of the Polyakov loops) and $\Delta$ is the two-dimensional Laplace operator $\Delta=\partial^2/ \partial t^2+\partial^2/\partial r^2$. As we are interested in evaluating the contribution for Polyakov-loop correlators, we assume that the $X$ field obeys periodic boundary conditions in the Euclidean time direction, $X(t,r)=X(t+N_t,r)$, and fixed boundary conditions in the direction of the spatial separation between the loops, $X(t,0)=X(t,R)=0$.
The interquark potential is defined as
\begin{equation}
\label{interquark_potential}
V(R)= - \lim_{N_t \to \infty } \frac{1}{N_t} \ln \left\{\int[D X] e^{-S[X]} \right\},
\end{equation}
where the functional integral on the right-hand side is performed over string worldsheet configurations satisfying the boundary conditions defined above. 

The Gaussian part of the action can be rewritten as
\begin{equation}
S=\sigma\int\limits_{0}^{N_t}dt\int\limits_{0}^{R}dr\left[1+
\frac{1}{2}X\left(1-\frac{2\alpha}{\sigma}\Delta\right)
\left(-\Delta\right)X\right].
\end{equation}
Carrying out the Gaussian integration, one obtains\footnote{Note that for regularized determinants in general it is not true that $\det(AB)=(\det A)(\det B)$, and one has to face the possible presence of multiplicative anomalies. However it can be shown that for Lapalace-type operators in two dimensions the anomaly vanishes and the above relation holds unchanged~\cite{Elizalde:1997nd}.}
\begin{equation}
\label{pot_from_Gaussian}
V(R)=\lim_{N_t \to \infty} \left\{ \sigma R+\frac{1}{2 N_t} \tr \ln(-\Delta)+ \frac{1}{2 N_t} \tr \ln\left(1-\frac{\Delta}{m^2}\right) \right\}, \qquad \mbox{with} \;\;
m^2 = \frac{\sigma}{2 \alpha}.
\end{equation}
The parameter $m$, with dimensions of a mass, encodes the contribution due to the extrinsic curvature.

Eq.~(\ref{pot_from_Gaussian}) reveals that, at the Gaussian level, the interquark potential is the sum of a contribution from a free massless bosonic field plus a free massive bosonic field. The mass of the latter is $m$, and is inversely proportional to the square root of the (dimensionless) coefficient of the extrinsic-curvature term appearing in the action. Following Polyakov's analysis, we may assume that in the 3D $\U(1)$ case $m$ is proportional to the mass of the lightest glueball in the theory, $m_0$.

The operator traces appearing in eq.~(\ref{pot_from_Gaussian}) can be readily evaluated using a zeta-function regularization, leading to the standard L\"uscher term for the contribution from the massless term and to the following contribution for the massive case~\cite{Nesterenko:1997ku, Klassen:1990dx}:
\begin{equation}
V(R)= \sigma R + \VNG(R) + \Vext(R,m),
\end{equation}
where $\VNG(R)$ and $\Vext(R,m)$ are the Gaussian limits of the Nambu-Goto and of the extrinsic-curvature contributions respectively:
\begin{eqnarray}
&& \VNG(R)\equiv \lim_{N_t\to\infty}\frac{1}{2 N_t}\mbox{Tr} \ln(-\Delta) 
=-\frac{\pi}{24 R}, \\
&& \Vext(R,m)\equiv\lim_{N_t\to\infty} \frac{1}{2 N_t}
\mbox{Tr}
\ln\left(1-\frac{\Delta}{m^2}\right)
=-\frac{m}{2\pi}\sum\limits_{n=1}^{
\infty} \frac{K_1 \left(2n m R\right) }{n},
\end{eqnarray}
where $K_\alpha(z)$ denotes a modified Bessel function of the second kind. 

$\Vext(R,m)$ has very interesting analytical properties. It is an analytic function of $R$ and $m$  for real positive values of $mR$. It has a logarithmic branching point at $R=0$ and, what is most interesting for our purposes, a set of square-root singularities for negative values of $(mR)^2$. The first of these singularities is located at $(mR)^2=-\pi^2$, and defines the radius of convergence of the expansion of the function in terms of $mR$. As we will show in sect.~\ref{sect:results}, most of our data are below this threshold. Using the Taylor expansion of modified Bessel functions and the $\zeta$-function regularization for the infinite sums, for $0 < mR < \pi$ one finds 
\begin{equation}
\label{expansion}
\Vext(R,m)= -\frac{\pi}{24 R} + \frac{m}{4} + \frac{m^2R}{4\pi}\left[\ln{\left(\frac{mR}{2\pi}\right)} +\gammaE -\frac12 \right]
+\frac{m^2R}{2\pi} \sum_{n=1}^{\infty}\frac{\Gamma\left(\frac32\right)\zeta(2n+1)}{\Gamma(n+2)\Gamma\left(n-\frac12\right)}
\left(\frac{mR}{\pi}\right)^{2n},
\end{equation}
where $\gammaE=0.5772156649\dots$ is the Euler-Mascheroni constant and $\zeta(x)$ denotes the Riemann zeta function.

A few comments may be useful to better understand this result.
\begin{itemize}
\item As we mentioned above, $\Vext(R,m)$ can be interpreted as a massive perturbation of the $c=1$ free bosonic theory. In fact, the combination 
\begin{equation}
c_0(2mR) = -\frac{24 R}{\pi} \Vext(R,m)
\end{equation}
coincides with the ground state scaling function $c_0(mR)$ introduced in ref.~\cite{Klassen:1990dx} to describe this perturbation. As expected, $c_0(mR)$ is a monotonically decreasing function of its argument and interpolates between $1$ for $mR=0$ and $0$ for $mR \to \infty$. In this respect it is interesting to notice the analogy with the Nambu-Goto case: while the Nambu-Goto model can be described as an irrelevant \emph{massless} perturbation of the $c=1$ free bosonic conformal field theory (CFT) in two dimensions~\cite{Dubovsky:2012sh, Caselle:2013dra}, the rigid string is described by a relevant \emph{massive} perturbation of the same CFT. 
\item The presence of a massive degree of freedom on the worldsheet of the confining string has been recently proposed as a way to explain the deviations from the expected Nambu-Goto behavior observed in Monte Carlo simulations of $\SU(N)$ Yang-Mills theories~\cite{Dubovsky:2013gi, Dubovsky:2014fma}. Our results can be considered as an explicit realization of this proposal in the 3D $\U(1)$ model.
\item The expansion on the right-hand side of eq.~(\ref{expansion}) agrees with the result for the small-$R$ regime obtained by Braaten, Pisarski and Tse in the $D\to\infty$ limit in ref.~\cite{Braaten:1986bz}. This shows that, in this regime, their result also holds for finite $D$.
\item In the $mR \to 0$ limit, the free bosonic theory is recovered: thus we find a second ``L\"uscher'' term, in addition to the one from $\VNG(R)$. As long as $m$ is small (in particular for $m < \pi \sqrt{\sigma}$), we should thus expect a major effect of the extrinsic-curvature term in the finite-size correction to the interquark potential. If $m$ is proportional to $m_0$ as suggested by Polyakov, see eq.~(\ref{Polyakov}), then, due to eq.~(\ref{ratio_m0}), the contribution of the extrinsic curvature should become more and more important as $\beta$ increases, becoming dominant in the continuum limit.
\end{itemize}
In the large-$R$ limit, $\Vext(R,m)$ decreases exponentially. Its behavior is dominated by the lowest-index Bessel function appearing in the sum:
\begin{equation}
\Vext(R,m) \simeq -\sqrt{\frac{m}{16\pi R}} e^{-2mR} \qquad \mbox{for} \;\; R \gg \frac{1}{m} .
\end{equation}
This is the typical behavior expected for a massive perturbation of a CFT in two dimensions, and agrees with well-known results on how the extrinsic-curvature coupling for the rigid string varies under renormalization-group transformations~\cite{Polyakov:1986cs, Kleinert:1986bk}. As a consequence, the contribution of $\Vext(R,m)$ becomes negligible in the infrared limit at fixed $m$ (but it can remain finite in the infrared limit if the latter is taken at fixed $mR$).

\subsection{Higher-order corrections}
\label{subsect:higherorder}

The contribution to the interquark potential due to the boundary term in eq.~(\ref{bounda2}) and to the next-to-leading-order term in the Nambu-Goto action can be evaluated perturbatively, taking into account both the contribution due to the Nambu-Goto action, and to the extrinsic-curvature term in the Gaussian integration~\cite{German:1989vk}. For the boundary term, the extrinsic-curvature term leads to contributions beyond our resolution, hence we are left with the standard free bosonic result derived in refs.~\cite{Aharony:2010cx, Billo:2012da}. Its contribution to the interquark potential in the large-$N_t$ limit is
\begin{equation}
\Vb=-b_2\frac{\pi^3}{60 R^4}.
\label{nextboundary}
\end{equation}
For the next-to-leading-order contribution of the Nambu-Goto action, the situation is slightly more complicated, and involves the usual Nambu-Goto correction, $O(R^{-3})$, and an additional, $O(R^{-4})$ term, in principle detectable in our simulations. The former contribution reads~\cite{Arvis:1983fp}
\begin{equation}
V_{1}=-\left(\frac{\pi}{24}\right)^2\frac{1}{2\sigma R^3}.
\label{next}
\end{equation}
On the other hand, the latter was computed in the large-$D$ limit in ref.~\cite{Braaten:1986bz}:
\begin{equation}
V_{2}=-\left(\frac{\pi D}{24}\right)^2\frac{3}{20 m\sigma R^4},
\end{equation}
but its expression for generic $D$, evaluated later in ref.~\cite{German:1989vk}, turns out to be affected by large finite-$D$ corrections:
\begin{equation}
V_{2}'=-(D-2)(D-10)\left(\frac{\pi}{24}\right)^2\frac{3}{20 m\sigma R^4} = V_{2} \cdot \left( 1- \frac{12}{D} + \frac{20}{D^2} \right).
\label{german}
\end{equation}
While it would be interesting to test this finite-$D$ dependence numerically, unfortunately out present numerical results do not allow us to disentangle this correction from the contribution due to the boundary term, with the same $1/R^4$ dependence. However, this could be possible in the future, with precise simulations on a wider range of lattice spacings, thanks to the different scaling behavior of the two terms.

\section{Numerical results}
\label{sect:results}

We carried out a set of simulations of the $\U(1)$ lattice model in its dual formulation, combining a conventional Metropolis algorithm~\cite{Metropolis:1953am} with the snake algorithm~\cite{deForcrand:2000fi} and with hierarchical lattice updates~\cite{Caselle:2002ah}. The simulations were performed on cubic lattices of size $L^2\times N_t$ ranging from $L=N_t=64a$ to $L=N_t=128a$, for five values of the Wilson lattice parameter from $\beta=1.7$ to $\beta=2.4$. These values were chosen in order to access a sufficiently wide range of values for $\sigma$ and $m_0$. To avoid systematic finite-volume effects, we always chose $L$ in such a way that $L>10/\sqrt{\sigma}$ and $L>10/m_0$. Details on the simulation settings are reported in tab.~\ref{tab:simul}. 

In our simulations we evaluated the ratio between two-point Polyakov-loop correlators at distances differing by one lattice spacing,
\begin{equation}
Q(R) = -\frac{1}{N_t}\ln{\frac{G(R+a)}{G(R)}}
\end{equation}
where $G(R)=\langle P^\star(x) P(x+R) \rangle$ and $P(x)$ denotes the Polyakov loop at the space site $x$. Note that, since $N_t$ has the dimensions of a length, $Q$ has energy dimension $1$. Using the numerical techniques discussed above, this quantity could be evaluated to high precision for several values of $\beta$ in the range $1/\sqrt{\sigma}<R<L/2$.
\begin{table}[ht]
\centering
\begin{tabular}{|c|c|c|c|}
\hline
$\beta$  &$\sigma a^2$  & $m_0 a$ & $L/a=N_t/a$\\
\hline
$1.7$ &$ 0.122764(2)$  & $0.88(1)$ & $64$\\
$1.9$ &$ 0.066824(6)$  & $0.56(1)$ & $64$\\
$2.0$ &$ 0.049364(2)$  & $0.44(1)$& $64$\\
$2.2$ &$ 0.027322(2)$  & $0.27(1)$ & $64$\\
$2.4$ &$ 0.015456(7)$  &$0.197(10)$& $128$\\
\hline
\end{tabular}
\caption{Information on the setup of our simulations.}
\label{tab:simul}
\end{table}

\subsection{Large deviations from the Nambu-Goto effective string predictions \dots}

We first tried to fit our numerical results with the standard Nambu-Goto effective string expectation, which for $Q(R)$ is\footnote{Our fits were carried out using the NG expression to all orders in the $1/R$  expansion, as in eq.~(\ref{NGfit1}). Within the precision of our data, fits obtained truncating the series to the $O(R^{-3})$ term give completely compatible results.}
\begin{equation}
\QNG(R) = \sigma \left[ \sqrt{(R+a)^2 - \frac{\pi}{12 \sigma}} - \sqrt{R^2 - \frac{\pi}{12 \sigma}}\,\right].
\label{NGfit1}
\end{equation}
We fitted  $Q(R)$ for $R$ ranging from $\Rmin$ to $R=L/2$, using the string tension $\sigma$ as the only fitting parameter. We started from $\Rmin \simeq 1/\sqrt{\sigma}$ and increased $\Rmin$ until we reached a value of the reduced $\chi^2$ close to $1$. The best-fit values for $\sigma$ obtained in this way are reported in the second column of tab.~\ref{tab:simul}. While for $\beta<2$ a $\redchisq$ of order one could be reached after a few lattice spacings, for $\beta \geq 2$ we had to choose larger and larger values of $\Rmin$. As an example, in the first three columns of tab.~\ref{tab:fit2.2} we report the fit results in the case of $\beta=2.2$, where a $\redchisq \lesssim 1$ (and a corresponding plateau in the best-fit value for $\sigma$) could only be reached for $\Rmin=26a$, which corresponds to $\Rmin\sqrt{\sigma}=4.3$. The magnitude of these deviations can be appreciated looking at fig.~\ref{fig:fig2}, where we plotted the $[Q(R)-\QNG(R)]$ differences, using for $\QNG(R)$ the asymptotic values of $\sigma$ reported in tab.~\ref{tab:simul}. These numbers can be compared with analogous fit results in the 3D $\Z_2$ gauge theory~\cite{Caselle:2004jq}: in particular, our data for $\beta=2.2$ and $\beta=2.4$ can be compared with the data reported in ref.~\cite{Caselle:2004jq} for the $\beta=0.75180$ sample of the 3D $\Z_2$ gauge theory, for which we had $\sigma a^2=0.010532(4)$, $L=80a$ and a similar level of precision for the $Q(R)$ values. In the 3D $\Z_2$ gauge theory we could fit the data with the L\"uscher correction \emph{alone}, already starting from $R\sqrt{\sigma}=1.8$, and the difference between the data and the L\"uscher correction was almost completely accounted for by the $1/R^3$ term of the Nambu-Goto action, see eq.~(\ref{next}). It was only by further improving the data precision and using sophisticated simulation methods, that deviations from the Nambu-Goto predictions, beyond the $1/R^3$ order, could be observed in the 3D $\Z_2$ model~\cite{Caselle:2007yc, Caselle:2010pf}.

\begin{sidewaystable}
\begin{tabular}{|c|c|c|c|c|c|c|c|c|c|c|c|}
\hline
$\Rmin\sqrt{\sigma}$ & \multicolumn{3}{|c|}{NG} &
\multicolumn{4}{|c|}{NG $+\Vext$} & \multicolumn{4}{|c|}{NG $+\Vext + V_2'$}  \\
\hline
 & d.o.f. & $\sigma a^2$ & $\redchisq$ & d.o.f. & $\sigma a^2$ & $ma$ &
$\redchisq$ & d.o.f. & $\sigma a^2$  & $ma$ & $\redchisq$ \\
\hline
$ 0.99 $ & $ 26 $ & $ 0.027428(1) $ & $ 883.2 $ & $ 25 $ & $ 0.027376(1) $ & $ 0.2199(8) $ & $ 284.76 $ & $ 25 $ & $ 0.027378(1) $ & $ 0.1252(9) $ & $ 401.86 $\\
$ 1.16 $ & $ 25 $ & $ 0.027416(1) $ & $ 714.14 $ & $ 24 $ & $ 0.027353(1) $ & $ 0.1681(7) $ & $ 96.45 $ & $ 24 $ & $ 0.027336(2) $ & $ 0.1013(9) $ & $ 90.59 $\\
$ 1.32 $ & $ 24 $ & $ 0.027402(1) $ & $ 513.65 $ & $ 23 $ & $ 0.027340(1) $ & $ 0.1444(8) $ & $ 33.5 $ & $ 23 $ & $ 0.027317(2) $ & $ 0.0934(9) $ & $ 15.87 $\\
$ 1.49 $ & $ 23 $ & $ 0.027391(1) $ & $ 357.44 $ & $ 22 $ & $ 0.027333(1) $ & $ 0.1322(9) $ & $ 13.8 $ & $ 22 $ & $ 0.027315(2) $ & $ 0.0944(9) $ & $ 2.71 $\\
$ 1.65 $ & $ 22 $ & $ 0.027381(1) $ & $ 259.13 $ & $ 21 $ & $ 0.027328(2) $ & $ 0.1250(9) $ & $ 6.38 $ & $ 21 $ & $ 0.027316(2) $ & $ 0.0959(10) $ & $ 1.26 $\\
$ 1.82 $ & $ 21 $ & $ 0.027374(1) $ & $ 200.34 $ & $ 20 $ & $ 0.027326(2) $ & $ 0.121(1) $ & $ 3.83 $ & $ 20 $ & $ 0.027317(2) $ & $ 0.097(1) $ & $ 1.2 $\\
$ 1.98 $ & $ 20 $ & $ 0.027362(1) $ & $ 115.51 $ & $ 19 $ & $ 0.027323(2) $ & $ 0.116(1) $ & $ 2.17 $ & $ 19 $ & $ 0.027317(2) $ & $ 0.097(1) $ & $ 1.25 $\\
$ 2.15 $ & $ 19 $ & $ 0.027354(1) $ & $ 73.22 $ & $ 18 $ & $ 0.027321(2) $ & $ 0.111(2) $ & $ 1.04 $ & $ 18 $ & $ 0.027316(2) $ & $ 0.096(1) $ & $ 1.26 $\\
$ 2.31 $ & $ 18 $ & $ 0.027347(1) $ & $ 43.35 $ & $ 17 $ & $ 0.027321(2) $ & $ 0.111(2) $ & $ 1.1 $ & $ 17 $ & $ 0.027318(2) $ & $ 0.098(2) $ & $ 1.06 $\\
$ 2.48 $ & $ 17 $ & $ 0.027342(1) $ & $ 26.71 $ & $ 16 $ & $ 0.027321(2) $ & $ 0.111(2) $ & $ 1.16 $ & $ 16 $ & $ 0.027318(2) $ & $ 0.099(2) $ & $ 1.08 $\\
$ 2.64 $ & $ 16 $ & $ 0.027337(1) $ & $ 12.07 $ & $ 15 $ & $ 0.027320(2) $ & $ 0.109(3) $ & $ 1.2 $ & $ 15 $ & $ 0.027319(2) $ & $ 0.100(3) $ & $ 1.15 $\\
$ 2.81 $ & $ 15 $ & $ 0.027336(1) $ & $ 9.3 $ & $ 14 $ & $ 0.027319(2) $ & $ 0.107(4) $ & $ 1.23 $ & $ 14 $ & $ 0.027318(3) $ & $ 0.099(3) $ & $ 1.22 $\\
$ 2.98 $ & $ 14 $ & $ 0.027334(1) $ & $ 6.3 $ & $ 13 $ & $ 0.027319(3) $ & $ 0.107(4) $ & $ 1.32 $ & $ 13 $ & $ 0.027319(3) $ & $ 0.100(4) $ & $ 1.3 $\\
$ 3.14 $ & $ 13 $ & $ 0.027334(1) $ & $ 6.5 $ & $ 12 $ & $ 0.027319(3) $ & $ 0.107(5) $ & $ 1.42 $ & $ 12 $ & $ 0.027319(3) $ & $ 0.100(4) $ & $ 1.4 $\\
$ 3.31 $ & $ 12 $ & $ 0.027332(1) $ & $ 4.92 $ & $ 11 $ & $ 0.027319(3) $ & $ 0.106(5) $ & $ 1.54 $ & $ 11 $ & $ 0.027319(3) $ & $ 0.100(5) $ & $ 1.52 $\\
$ 3.47 $ & $ 11 $ & $ 0.027329(2) $ & $ 2.92 $ & $ 10 $ & $ 0.027320(3) $ & $ 0.111(8) $ & $ 1.6 $ & $ 10 $ & $ 0.027320(3) $ & $ 0.104(7) $ & $ 1.57 $\\
$ 3.64 $ & $ 10 $ & $ 0.027329(2) $ & $ 3.16 $ & $ 9 $ & $ 0.027316(4) $ & $ 0.096(7) $ & $ 0.92 $ & $ 9 $ & $ 0.027315(4) $ & $ 0.091(7) $ & $ 0.91 $\\
$ 3.8 $ & $ 9 $ & $ 0.027327(2) $ & $ 2.67 $ & $ 8 $ & $ 0.027314(5) $ & $ 0.092(8) $ & $ 0.97 $ & $ 8 $ & $ 0.027314(5) $ & $ 0.087(8) $ & $ 0.96 $\\
$ 3.97 $ & $ 8 $ & $ 0.027326(2) $ & $ 2.04 $ & $ 7 $ & $ 0.027314(5) $ & $ 0.09(1) $ & $ 1.11 $ & $ 7 $ & $ 0.027314(6) $ & $ 0.09(1) $ & $ 1.1 $\\
$ 4.13 $ & $ 7 $ & $ 0.027325(2) $ & $ 2.23 $ & $ 6 $ & $ 0.027308(7) $ & $ 0.08(1) $ & $ 0.85 $ & $ 6 $ & $ 0.027307(8) $ & $ 0.08(1) $ & $ 0.85 $\\
$ 4.3 $ & $ 6 $ & $ 0.027322(2) $ & $ 0.91 $ & $ 5 $ & $ 0.027314(8) $ & $ 0.09(2) $ & $ 0.69 $ & $ 5 $ & $ 0.027314(8) $ & $ 0.09(2) $ & $ 0.69 $\\
$ 4.46 $ & $ 5 $ & $ 0.027321(2) $ & $ 0.85 $ & $ 4 $ & $ 0.027315(10) $ & $ 0.10(3) $ & $ 0.87 $ & $ 4 $ & $ 0.02731(1) $ & $ 0.09(3) $ & $ 0.86 $\\
$ 4.63 $ & $ 4 $ & $ 0.027322(3) $ & $ 1.06 $ & $ 3 $ & $ 0.02730(2) $ & $ 0.07(2) $ & $ 0.74 $ & $ 3 $ & $ 0.02730(2) $ & $ 0.07(2) $ & $ 0.74 $\\
$ 4.79 $ & $ 3 $ & $ 0.027319(3) $ & $ 0.58 $ & $ 2 $ & $ 0.02731(2) $ & $ 0.10(8) $ & $ 0.82 $ & $ 2 $ & $ 0.02731(2) $ & $ 0.09(8) $ & $ 0.82 $\\
$ 4.96 $ & $ 2 $ & $ 0.027317(4) $ & $ 0.6 $ & $ - $ & $ - $ & $ - $ & $ - $ & $ 1 $ & $ 0.02730(2) $ & $ - $ & $ 1.18 $\\

\hline
\end{tabular}
\caption{Fits of our numerical results at $\beta=2.2$ to various effective string \emph{Ans\"atze}. In the third and fourth column we report the results of the fit of $Q(R)$ with the Nambu-Goto \emph{Ansatz} (i.e. using $\QNG(R)$). The remaining columns show the results of the two-parameter fits using the Nambu-Goto + rigid-string \emph{Ansatz}: first assuming only the Gaussian term for the rigid string (i.e. using $[\QNG(R)+\Qr(R)]$) and then (last three columns) using also the next-to-leading-order rigid-string correction (i.e. using $[\QNG(R)+\Qr'(R)]$).}
\label{tab:fit2.2}
\end{sidewaystable}

\subsection{\dots that cannot be fitted by a boundary correction}

Next, we tested if the deviations from the prediction of the Nambu-Goto action could be fitted by a boundary correction of the type in eq.~(\ref{nextboundary}). To this end, following the notation used in refs.~\cite{Aharony:2010cx, Billo:2012da}, we fitted the $[Q(R)-\QNG(R)]$ differences to the boundary correction
\begin{equation}
\Qb(R) = - \frac{b_2\pi^3}{60}\left[ \left( \frac{1}{R+a} \right)^4 - \left( \frac{1}{R} \right)^4 \right].
\label{Qb_with_b2_coefficient}
\end{equation}
Note that $b_2$ has energy dimension $-3$. We found, as in the case of the previous fits, that reasonable $\redchisq$ values could only be reached for very large values of $\Rmin$. Even more important, the best-fit values for $b_2$ thus obtained did not show the expected scaling behavior. When the fits to eq.~(\ref{Qb_with_b2_coefficient}) are carried out expressing all quantities in the appropriate lattice units (i.e. using $a\Qb$ instead of $\Qb$, $R/a$ instead of $R$, et c.), one extracts results for the dimensionless ratio $b_2/a^3$. If $b_2$ is a physical (i.e. non-renormalized) quantity in the continuum limit, then the $b_2/a^3$ values obtained from the fit should scale as $a^{-3}$, or, equivalently, as $(\sigma a^2)^{-3/2}$. Instead, we found that the values of $(b_2/a^3)/(\sigma a^2)^{-3/2}=b_2\sigma^{3/2}$ obtained from the fits range from $b_2\sigma^{3/2}=0.033(3)$ for $\beta=1.7$, up to $b_2\sigma^{3/2}=0.62(6)$ for $\beta=2.4$. The fact that the $b_2$ parameter increases as a function of $\beta$ agrees with the observation pointed out above, that the deviations from the Nambu-Goto action become larger and larger as $\beta$ increases.

As a complementary test, we also performed a two-parameter fit of the $[Q(R)-\QNG(R)]$ differences to a correction term with a free exponent $b$,
\begin{equation}
\label{k_and_b_fit}
\Qb'(R) = k \left[ \left( \frac{a}{R+a} \right)^b - \left( \frac{a}{R} \right)^b \right],
\end{equation}
where $k$ has energy dimension $1$. At all lattice spacings, we found values of the exponent ranging between $2$ and $3$, and thus incompatible with a boundary-type correction. Moreover, reasonable values of $\redchisq$ could only be reached for very large values of $\Rmin$, for which the coefficient $k$ was almost compatible with zero, within its uncertainties. As an example, the results of these fits for $\beta=2.2$ are shown in tab.~\ref{tab:fit-power-2.2}.
\begin{table}[ht]
 \centering
\begin{tabular}{|c|c|c|c|c|}
\hline
$\Rmin\sqrt{\sigma}$ & d.o.f.  & $ak$ &  $b$ & $\redchisq$ \\
\hline
$ 0.99 $ & $ 25 $ & $ -0.0394(9) $ & $ 1.16(2) $ & $ 73.27 $\\
$ 1.16 $ & $ 24 $ & $ -0.087(4) $ & $ 1.6(3) $ & $ 26.66 $\\
$ 1.32 $ & $ 23 $ & $ -0.17(1) $ & $ 1.95(4) $ & $ 12.84 $\\
$ 1.49 $ & $ 22 $ & $ -0.33(3) $ & $ 2.24(5) $ & $ 8.12 $\\
$ 1.65 $ & $ 21 $ & $ -0.63(9) $ & $ 2.54(7) $ & $ 5.39 $\\
$ 1.82 $ & $ 20 $ & $ -1.1(2) $ & $ 2.78(9) $ & $ 4.01 $\\
$ 1.98 $ & $ 19 $ & $ -2.5(7) $ & $ 3.1(1) $ & $ 2.91 $\\
$ 2.15 $ & $ 18 $ & $ -11(5) $ & $ 3.7(2) $ & $ 1.38 $\\
$ 2.31 $ & $ 17 $ & $ -17(11) $ & $ 3.9(2) $ & $ 1.36 $\\
$ 2.48 $ & $ 16 $ & $ -27(21) $ & $ 4.0(3) $ & $ 1.39 $\\
$ 2.64 $ & $ 15 $ & $ -60(80)$ & $ 4.3(5) $ & $ 1.42 $\\
\hline
\end{tabular}
\caption{Results of the two-parameter fits of the $[Q(R)-\QNG(R)]$ differences at $\beta=2.2$ to the $k \cdot \left[ (1+R/a)^{-b}-(R/a)^{-b} \right]$ functional form, defined in eq.~(\protect\ref{k_and_b_fit}).}
\label{tab:fit-power-2.2}
\end{table}

\subsection{Fit of the data with a rigid-string \emph{Ansatz}}

Much better fits were obtained by fitting the $[Q(R)-\QNG(R)]$ differences with the rigid-string prediction, i.e. with
\begin{equation}
\Qr(R) = -\frac{m}{2\pi}\sum_{n=1}^{\infty} \frac{K_1\left(2 n m (R+a)\right)-K_1(2 n m R)}{n},\,\,\,\, m=\sqrt{\frac{\sigma}{2\alpha}}.
\end{equation}
In practice, we truncated the sum over Bessel functions at $n=100$ and verified that for all values of $R$ and $\beta$ this gave differences well below the statistical uncertainties of our data.

Carrying out one-parameter fits with $m$ as the only free parameter, we could successfully fit the data with much smaller values of $\Rmin$, and the resulting values of $m$ had the expected scaling behavior, proportional to the glueball mass $m_0$ (see tab.~\ref{tab:fit2}). 

\begin{table}[ht]
 \centering
 \begin{tabular}{|c|c|c|c|}
 \hline
 $\beta$ & $m a$ & $m_0 a$ & $m/m_0$\\
 \hline
 $1.7$ & $0.28(9)$ & $0.88(1)$ & $0.32(10)$\\
 $1.9$ & $0.25(4)$ & $0.56(1)$ & $0.45(7)$\\
 $2.0$ & $0.17(2)$ & $0.44(1)$& $0.39(4)$\\
 $2.2$ & $0.11(1)$ & $0.27(1)$ & $0.41(4)$\\
 $2.4$ & $0.06(2)$ &$0.20(1)$& $0.30(10)$\\
 \hline
 \end{tabular}
 \caption{Best-fit results for $m$ obtained using a three-parameter fit to our data, as explained in the text.}
 \label{tab:fit2}
\end{table}

Since the $m_0/\sqrt{\sigma}$ ratio in the $\U(1)$ lattice model is not constant, but rather is expected to scale according to eq.~(\ref{ratio_m0}), this explains why these corrections become more and more important as $\beta$ increases. 

\begin{figure}[-t]
\centerline{\includegraphics[width=\textwidth]{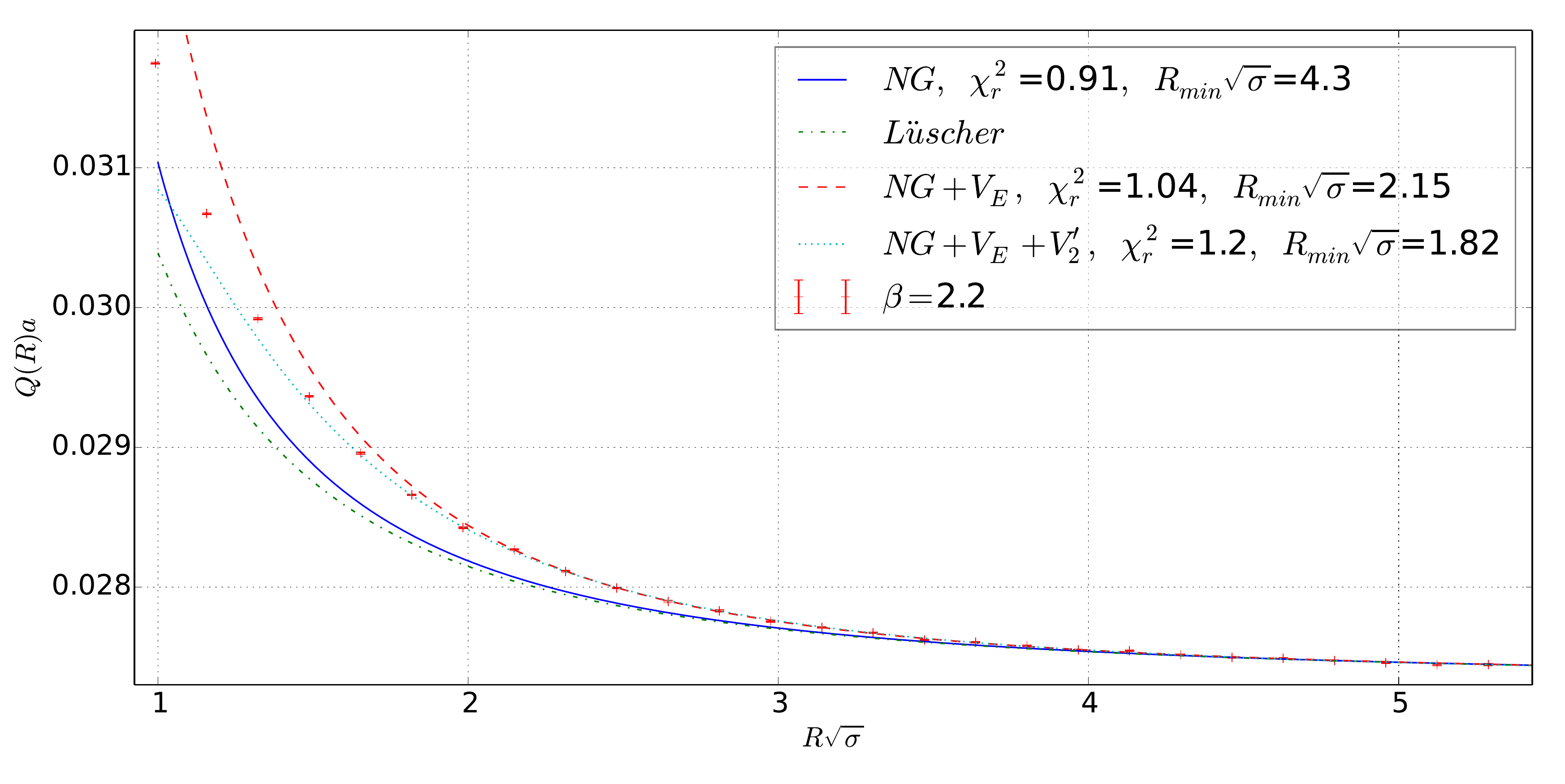}}
\caption{Numerical data (at $\beta=2.2$) and best-fit curves for the functional forms corresponding to $\QNG$, $\QNG+$ L\"uscher,$\Vext$, and $\Vext+V'_2$, as defined in the text. The last two are two-parameter ($\sigma$ and $m$) fits.\label{fig:fig1}}
\end{figure}

\begin{figure}[-t]
\centerline{\includegraphics[width=\textwidth]{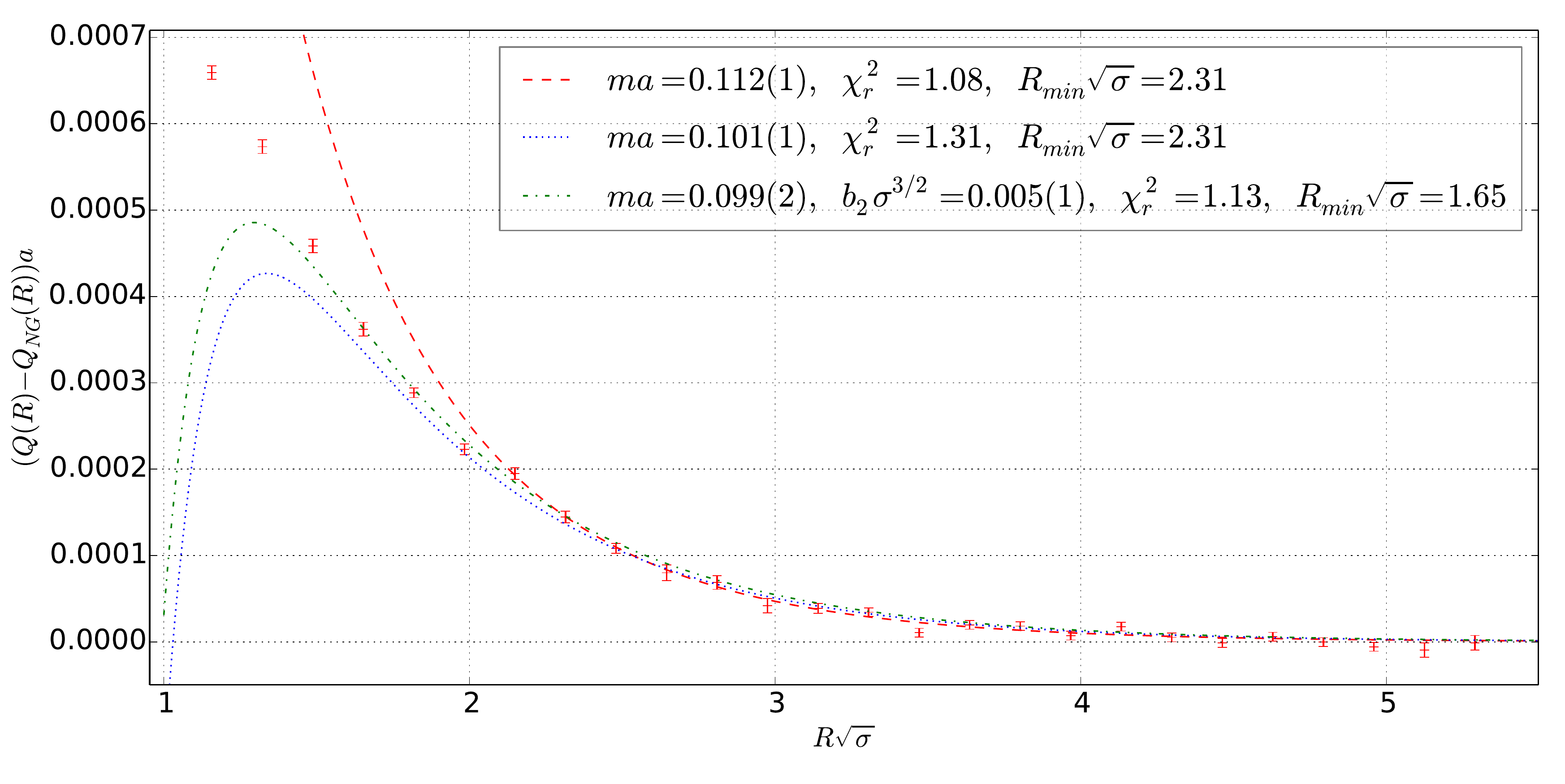}}
\caption{Fits of the $[Q-\QNG(R)]$ differences at $\beta=2.2$, with $\Vext$, $\Vext+V'_2$ and $\Vext+V'_2+$ boundary term.\label{fig:fig2}}
\end{figure}

Tab.~\ref{tab:res2.2} shows an example of our results for $\beta=2.2$: a $\redchisq$ of order one could be reached already for $\Rmin\sqrt{\sigma}=2.15$, which is a remarkable improvement over the one-parameter fit to the pure Nambu-Goto prediction. Our data for $Q(R)$ at $\beta=2.2$ are plotted in fig.~\ref{fig:fig1}, together with with the fit results.

\begin{table}
\centering{
\begin{tabular}{|c|c|c|c|c|c|c|c|c|}
\hline
$\Rmin\sqrt{\sigma}$ & &\multicolumn{2}{|c|}{$\Vext$} & \multicolumn{2}{|c|}{$\Vext + V_2'$} & \multicolumn{3}{|c|}{$\Vb$} \\
\hline
& d.o.f. & $ma$ & $\redchisq$ & $m a$ & $\redchisq$ & d.o.f. & $b_2 \sigma^{3/2}$ & $\redchisq$ \\
\hline
$ 0.99 $ & $ 26 $ & $ 0.2054(7) $ & $ 340.17 $ & $ 0.1182(7) $ & $ 462.7 $ &   $ 25 $ & $ 0.0198(2) $ & $ 14.21 $\\
$ 1.16 $ & $ 25 $ & $ 0.1596(6) $ & $ 113.36 $ & $ 0.0981(6) $ & $ 90.47 $ &   $ 24 $ & $ 0.0154(3) $ & $ 4.95 $\\
$ 1.32 $ & $ 24 $ & $ 0.1390(6) $ & $ 38.56 $ & $ 0.0952(6) $ & $ 15.64 $   &   $ 23 $ & $ 0.0114(6) $ & $ 2.46 $\\
$ 1.49 $ & $ 23 $ & $ 0.1287(7) $ & $ 15.32 $ & $ 0.0971(6) $ & $ 3.34 $   &   $ 22 $ & $ 0.0073(10) $ & $ 1.32 $\\                                                                                       
$ 1.65 $ & $ 22 $ & $ 0.1228(7) $ & $ 6.78 $ & $ 0.0984(7) $ & $ 1.76 $    &   $ 21 $ & $ 0.005(1) $ & $ 1.13 $\\                                                                                         
$ 1.82 $ & $ 21 $ & $ 0.1194(8) $ & $ 3.9 $ & $ 0.0990(7) $ & $ 1.64 $      &   $ 20 $ & $ 0.004(2) $ & $ 1.17 $\\                                                                                         
$ 1.98 $ & $ 20 $ & $ 0.1153(10) $ & $ 2.07 $ & $ 0.0997(9) $ & $ 1.62 $   &   $ 19 $ & $ 0.004(3) $ & $ 1.23 $\\                                                                                         
$ 2.15 $ & $ 19 $ & $ 0.112(1) $ & $ 1.03 $ & $ 0.100(1) $ & $ 1.7 $         &   $ 18 $ & $ 0.009(5) $ & $ 1.17 $\\                                                                                         
$ 2.31 $ & $ 18 $ & $ 0.112(1) $ & $ 1.08 $ & $ 0.101(1) $ & $ 1.31 $      &   $ 17 $ & $ - $ & $ 1.05 $\\                                                                               
$ 2.48 $ & $ 17 $ & $ 0.112(2) $ & $ 1.15 $ & $ 0.103(1) $ & $ 1.24 $      &   $ 16 $ & $ - $ & $ 1.05 $\\                                                                              
$ 2.64 $ & $ 16 $ & $ 0.112(2) $ & $ 1.21 $ & $ 0.104(2) $ & $ 1.25 $      &   $ 15 $ & $ - $ & $ 1.11 $\\                                                                              
$ 2.81 $ & $ 15 $ & $ 0.111(2) $ & $ 1.28 $ & $ 0.104(2) $ & $ 1.34 $      &   $ 14 $ & $ - $ & $ 1.18 $\\                                                                              
$ 2.98 $ & $ 14 $ & $ 0.112(3) $ & $ 1.35 $ & $ 0.105(2) $ & $ 1.36 $      &   $ 13 $ & $ - $ & $ 1.24 $\\                                                                              
$ 3.14 $ & $ 13 $ & $ 0.112(3) $ & $ 1.43 $ & $ 0.105(3) $ & $ 1.44 $      &   $ 12 $ & $ - $ & $ 1.25 $\\                                                                              
$ 3.31 $ & $ 12 $ & $ 0.113(3) $ & $ 1.54 $ & $ 0.106(3) $ & $ 1.54 $      &   $ 11 $ & $ - $ & $ 1.36 $\\                                                                             
$ 3.47 $ & $ 11 $ & $ 0.117(5) $ & $ 1.5 $ & $ 0.109(4) $ & $ 1.47 $       &   $ 10 $ & $ - $ & $ 1.4 $\\                                                                               
$ 3.64 $ & $ 10 $ & $ 0.110(5) $ & $ 1.19 $ & $ 0.104(4) $ & $ 1.18 $       &   $ 9 $ & $ - $ & $ 0.81 $\\                                                                                           
$ 3.8 $ & $ 9 $ & $ 0.111(6) $ & $ 1.31 $ & $ 0.105(5) $ & $ 1.31 $        &  $ 8 $ & $ - $ & $ 0.89 $\\                                                                                            
$ 3.97 $ & $ 8 $ & $ 0.115(9) $ & $ 1.35 $ & $ 0.108(7) $ & $ 1.33 $       &   $ 7 $ & $ - $ & $ 0.97 $\\                                                                                
$ 4.13 $ & $ 7 $ & $ 0.113(10) $ & $ 1.51 $ & $ 0.106(8) $ & $ 1.49 $      &   $ 6 $ & $ - $ & $ 0.97 $\\                                                                                           
$ 4.3 $ & $ 6 $ & $ 0.14(4) $ & $ 0.85 $ & $ 0.13(2) $ & $ 0.82 $          &  $ 5 $ & $ - $ & $ 0.52 $\\                                                                                 
$ 4.46 $ & $ 5 $ & $ -      $ & $ 0.89 $ & $ 0.15(10) $&$ 0.86$ &   $ 4 $ & $ - $ & $ 0.6 $\\                                                                                
$ 4.63 $ & $ 4 $ & $ -      $ & $ 1.08 $ & $ 0.13(6) $&$ 1.06$ &    $ 3 $ & $ - $ & $ 0.66 $\\                                                                               
$ 4.79 $ & $ 3 $ & $ -      $ & $ 1.02 $ & $ -       $ & $0.91$      &   $ 2 $ & $ -$ & $ 0.35 $\\                                                                               
$ 4.96 $ & $ 2 $ & $ -      $ & $ 1.52 $ & $ -       $ & $ 1.36 $ &                       $ 1 $ & $ - $ & $ 0.25 $\\

\hline
\end{tabular}
}
\caption{Fits of the $[Q(R)-\QNG(R)]$ differences for
$\beta=2.2$, with $\sigma$ fixed to the value $\sigma a^2 =0.027322(2)$ determined for $\Rmin=26 a$ with a rigid-string \emph{Ansatz}. In the third and fourth column we report the results of the fit using only the Gaussian correction $\Qr(R)$, while in the fifth and sixth columns we list the results obtained including the next-to-leading-order term, too, i.e. using $\Qr'(R)$. In the last two columns we show the fit results for the $[Q(R)-\QNG(R)-\Qr'(R)]$ differences (with the fixed values $\sigma a^2 =0.027322(2)$ and $ma=0.099(2)$) with a boundary correction $\Qb(R)$.}
\label{tab:res2.2}
\end{table}

In fig.~\ref{fig:fig2} we show, again for $\beta=2.2$, the $[Q(R)-\QNG(R)]$ differences, together with the best-fit results for the rigid-string correction with and without a boundary term. The figure clearly reveals the magnitude of the deviations from the Nambu-Goto predictions.

\subsection{Numerical evidence for terms $O(R^{-4})$}

Next, we tested whether the next-to-leading-order correction $V_2'$ discussed in subsect.~\ref{subsect:higherorder} could be detected, within the precision of our data. To this end, we constructed the combination
\begin{equation}
\Qr'(R) = \Qr(R) + \frac{21}{20 m \sigma}\left(\frac{\pi}{24}\right)^2\left[ \frac{1}{(R+a)^4}-\frac{1}{R^4} \right]
\end{equation}
and used it to fit the $[Q(R)-\QNG(R)]$ differences, using $m$ as the only free parameter. In the fifth and sixth columns of tab.~\ref{tab:res2.2} we show an example of our results, in the $\beta=2.2$ case. The $\redchisq$ values exhibit significant improvement, particularly for  $1<\Rmin\sqrt{\sigma}<2$, even though they are still larger than $1$. At the same time, one also finds a change in the best-fit values for $m$, which is larger than our statistical uncertainties. We conclude that, within the precision of our data, this term cannot be neglected.

Since terms of order $1/R^4$ are non-negligible within the precision of our data, it is important to take the possible presence of \emph{both} a boundary and a rigid-string correction in the data into account. To detect a possible boundary correction, we fixed the best-fit values for $\sigma$ and $m$ obtained in the previous fits, and performed a one-parameter fit of the $[Q(R)-\QNG-\Qr'(R)]$ differences to the boundary correction $\Vb(R)$, with $b_2$ as the only free parameter. An example of our results for this type of fit (at $\beta=2.2$) is reported in the last two columns of tab.~\ref{tab:res2.2}. We find much better values of $\redchisq$ also for small $\Rmin\sqrt{\sigma}$. $\redchisq$ values around one are reached already for $\Rmin\sqrt{\sigma}\sim 1.65$, with a small but non-vanishing value of the rescaled $b_2\sigma^{3/2}$ parameter: $b_2\sigma^{3/2}=0.005(1)$. This indicates that also this term is non-negligible, and, at the level of precision of our results, should be taken into account in the analysis.

\subsection{Determination of the rigid-string parameter $m$}

Looking at the $\Rmin$ dependence of the fits, it is possible to see that the rigid-string correction could also influence the determination of $\sigma$. In order to test the quantitative impact of this possibility, in addition to the one-parameter fits of $[Q(R)-\QNG(R)]$ described above, we decided to perform two-parameter fits of $Q(R)$ to the functions $\QNG$ and $\Qr(R)$ (or $\Qr'(R)$) using both $\sigma$ and $m$ as free parameters. The results of these fits, again in the $\beta=2.2$ case, are reported in tab.~\ref{tab:fit2.2} (first using $\Qr(R)$ and then using $\Qr'(R)$). We see that $\sigma$ has a sizeable effect on the value of $m$ and on its statistical uncertainty and that, also in this case, including the next-to-leading-order correction of the rigid string changes  the best-fit results for $m$.

Based on this analysis, we conclude that both the value of $\sigma$ and that of $b_2$ may influence our estimate of $m$. To take this systematic ambiguity into account, we decided to use as our best-fit estimates for $m$ the results of a three-parameter fit to the data, with $\sigma$, $m$ and $b_2$ as free parameters. The drawback of this choice is that the resulting values of $m$ are affected by rather large uncertainties (reflecting our ignorance on the actual values of $\sigma$ and $b_2$). It is likely that this uncertainty will decrease as other observables are included in the analysis, like Wilson loops or high-temperature correlators of Polyakov loops. We plan to address this issue in future work. We report these estimates for $m$ in tab.~\ref{tab:fit2}. In the last column we report the $m/m_0$ ratio, which, as anticipated, shows good scaling behavior.

Taking both statistical and systematic uncertainties (including scaling violations) of the above values into account, we quote the value
\begin{equation}
\label{rigidstringparameterresult}
\frac{m}{m_0}=0.35(10)
\end{equation}
as our tentative estimate for the rigid-string parameter.

\section{Concluding remarks}
\label{sect:conclusions}

The results of our lattice simulations show that in the 3D $\U(1)$ model, as $\beta$ increases towards the continuum limit, the interquark potential shows strong deviations from the expectations of a Nambu-Goto effective string model. These deviations are described well by the one-loop contribution of an extrinsic-curvature term in the effective string action. Polyakov's derivation of the effective string description for the $\U(1)$ model suggests to relate the parameter $m$, which controls the rigid-string contribution, to the mass of the lightest glueball $m_0$. Since in the $\U(1)$ model the ratio $m_0/\sqrt{\sigma}$ decreases exponentially with $\beta$, we expect the continuum limit of the model to be dominated by rigid-string behavior, which is very different from the Nambu-Goto one. In this sense, it is really ``a different kind of string'', as anticipated in the title of the present article. Thus, the 3D $\U(1)$ lattice model turns out to be a perfect laboratory to study the cross-over from a purely Nambu-Goto string at low $\beta$ to a purely rigid string at large $\beta$.

The main differences between the two types of strings, which could be used to identify their behavior, can be summarized as follows.
\begin{itemize}
\item The field density profile around the string is (almost) a Gaussian in the case of a Nambu-Goto string, while it decreases exponentially for the rigid string. This exponential defines a new scale, known as the London penetration length in condensed matter theory, and sometimes denoted as intrinsic width in confining gauge theories~\cite{Muller:2004vv, Vyas:2010wg, Caselle:2012rp}.
\item While in the Nambu-Goto case the string width increases logarithmically with the interquark distance at zero temperature~\cite{Luscher:1980iy} and linearly at high temperature~\cite{Allais:2008bk}, the intrinsic width of the rigid string is constant \cite{Vyas:2010wg, Caselle:2012rp}.
\item At very short distances the coefficient of the L\"uscher term is doubled.
\end{itemize}

Our findings may have important implications also for other confining theories, including some of interest for elementary particle physics (like the $\SU(3)$ theory in four spacetime dimensions) or for condensed matter physics (like the 3D Ising model). In comparing the 3D $\U(1)$ model with other confining theories, one should consider two features: the scaling behavior of the $m_0/\sqrt{\sigma}$ ratio and the presence of a contribution due to the extrinsic-curvature term in the interquark potential. As discussed at the end of section~\ref{sect:U1}, the relative weight of the Nambu-Goto and extrinsic-curvature terms in the effective string description depends on the $m_0/\sqrt{\sigma}$ ratio, and its non-trivial dependence on the lattice spacing makes the rigidity term \emph{dominant} in the continuum limit of the 3D $\U(1)$ model. Such behavior, however, appears to be non-generic: for example, in non-Abelian lattice gauge theories typically $m_0/\sqrt{\sigma}$ remains constant when the lattice spacing tends to zero (up to small discretization artifacts). As a consequence, rigid-string effects may be present in the infrared regime of $\SU(N)$ gauge theories---and could perhaps explain some fine deviations from the Nambu-Goto string, that have been observed in recent simulations~\cite{Athenodorou:2010cs, Caselle:2007yc, Caselle:2010pf}, as well as the London penetration term in the string width~\cite{Cea:2012qw, Cea:2014uja}---but there is no reason to expect them to be dominant.

Thinking about connections between the model discussed in the present work and other theories, an interesting 3D model, in which the Kramers-Wannier duality transformation that we used here can be applied in the presence of matter, is the $\Z_2$ gauge-Higgs model: for a discussion, see ref.~\cite{Gliozzi:2002pd} and references therein.

Finally, it is interesting to note the analogy of our results with those obtained in the past in Abelian Higgs models in \emph{four} spacetime dimensions. Also in that case, in a certain limit (the so-called ``London limit''), a suitable duality transformation allows one to derive a description in terms of an effective bosonic string model, whose action includes a Nambu-Goto term and a rigidity term, as discussed in refs.~\cite{Forster:1974ga, Gervais:1974db, Lee:1993ty, Orland:1994qt, Sato:1994vz, Akhmedov:1995mw}. These works present a nice realization of the dual superconductor scenario, which was proposed forty years ago by 't~Hooft~\cite{'tHooft:1975pu} and by Mandelstam~\cite{Mandelstam:1974pi}. More recently, a similar approach has also been investigated in non-Abelian models: for a review, see ref.~\cite{Kondo:2014sta}. These works are part of the research efforts to derive an analytical understanding of confinement in non-Abelian gauge theories, in terms of objects that can be studied semi-classically, for which there has been significant progress in the past few years~\cite{Konishi:2007dn, Poppitz:2012sw, Poppitz:2012nz}. The main difference between our model and the situation in Abelian Higgs models is that, as we mentioned above, in our case the continuum limit is dominated by the rigidity term, while, a priori, there is no reason to expect a similar behavior in the Abelian Higgs models. Studying the relative weight of and the interplay between the string tension and the coefficient of the extrinsic curvature term in Abelian Higgs models in four spacetime dimensions would be a very interesting task, but one which clearly lies beyond the scope of the present article, hence we leave its numerical investigation for the future.

\vskip1.0cm
\noindent{\bf Acknowledgements}\\
We thank B.~Lucini for useful discussions and in particular for pointing to us a few subtle issues on the performance of the hierarchical snake algorithm. We thank F.~Gliozzi, M.~Meineri and E.~Poppitz for useful discussions and suggestions. This work is supported by the Spanish MINECO (grant FPA2012-31686 and ``Centro de Excelencia Severo Ochoa'' programme grant SEV-2012-0249).

\appendix
\section{Reabsorbing the rigidity term into the Gaussian term}
\label{app:reabsorbing}
\renewcommand{\theequation}{A.\arabic{equation}}
\setcounter{equation}{0}

At the Gaussian level, the effective string action, including the rigidity term, is given by eq.~(\ref{Gaussian_part_of_action}), which can be rewritten as 
\begin{equation}
S=\sigma\int\limits_{0}^{N_t}\ud t\int\limits_{0}^{R}\ud r\left[1+\frac{1}{2}X \left( 1 - \frac{2\alpha}{\sigma} \Delta\right)  (-\Delta) X\right]
\end{equation}
by integrating by parts and regrouping like terms.

At leading order (in an expansion in $\alpha/\sigma$), the rigidity term can be reabsorbed by a field redefinition: setting
\begin{equation}
 X'\left(\xi_0, \xi_1\right) = \left( 1 - \frac{2\alpha}{\sigma} \Delta \right)^{1/2} X(\xi_0, \xi_1)
 \label{field_redefinition}
\end{equation}
one gets
\begin{eqnarray}
 \partial_\alpha X' \partial^\alpha X' &=& \partial_\alpha \left[ \left( 1 - \frac{2\alpha}{\sigma} \Delta \right)^{1/2} X \right] \partial^\alpha \left[ \left( 1 - \frac{2\alpha}{\sigma} \Delta \right)^{1/2} X \right] \nonumber \\
 &=&  \partial_\alpha X \partial^\alpha X - \frac{\alpha}{\sigma}\partial_\alpha X \Delta\partial^\alpha  X - \frac{\alpha}{\sigma} \Delta \partial_\alpha X \partial^\alpha X + O\left(\left(\alpha/\sigma\right)^2\right) \nonumber \\
 &=&  \partial_\alpha X \partial^\alpha X + \frac{\alpha}{\sigma} (\Delta X )^2 + O\left(\left(\alpha/\sigma\right)^2\right) .
\end{eqnarray}
The change on the functional measure induced by the field redefinition in eq.~(\ref{field_redefinition}) can be worked out as follows. Let $\lambda_{ab}$ and $\Phi_{ab}$ denote the eigenvalues and eigenfunctions of the $\Delta$ operator:
\begin{equation}
 \Delta \Phi_{ab} = -\lambda_{ab} \Phi_{ab}.
\end{equation}
Then, the $\Phi_{ab}$'s are also eigenfunctions of the $\left( 1-\frac{\alpha}{\sigma} \Delta\right)^{1/2}$ operator:
\begin{equation}
 \left( 1-\frac{2\alpha}{\sigma} \Delta\right)^{1/2} \Phi_{cd} = \omega_{cd}\Phi_{cd} = \left( 1 +  \frac{2\alpha}{\sigma} \lambda_{cd}\right)^{1/2} \Phi_{cd}.
\end{equation}
Expanding $X$ and $X'$ in a basis of eigenfunctions of $\Delta$,  eq.~(\ref{field_redefinition}) implies that
\begin{eqnarray}
 X(\xi_0, \xi_1) &=& \sum_{a,b} x_{ab} \Phi_{ab}(\xi_0, \xi_1), \\
 X'(\xi_0, \xi_1) &=& \sum_{c,d} x'_{cd} \Phi_{cd}(\xi_0, \xi_1)=\sum_{cd}x_{cd}\left(1+\frac{2\alpha}{\sigma}\lambda_{cd}\right)^{1/2} \Phi_{cd}.
\end{eqnarray}
As a consequence,
\begin{equation}
 [D X] = \prod_{ab} \ud x_{ab} = \prod_{cd} \frac{1}{\left( 1+\frac{2\alpha}{\sigma} \lambda_{cd}\right)^{1/2}} \prod_{ab} \ud x_{ab} \left( 1+\frac{2\alpha}{\sigma}
 \lambda_{ab}\right)^{1/2} = \left[ \det \left( 1-\frac{2\alpha}{\sigma} \Delta \right)^{1/2} \right]^{-1}_{R,N_t} [DX'],
\end{equation}
which implies
\begin{equation}
 \int [DX] \exp ( -S ) = \frac{\int [DX']  \exp \left\{ -\sigma\int\limits_{0}^{N_t}\ud t\int\limits_{0}^{R}\ud r\left[1+\frac{1}{2}\partial_\alpha X'\cdot\partial^\alpha
 X'\right]\right\} }{\left[ \det \left(1-\frac{2\alpha}{\sigma} \Delta \right)^{1/2}\right]_{R,N_t}}.
\end{equation}
Using this result in eq.~(\ref{interquark_potential}), one obtains
\begin{equation}
\label{sqrt}
 V(R,N_t)=  \sigma R + \frac{1}{2N_t} \tr\left[ \ln{\left(1-\frac{2\alpha}{\sigma} \Delta \right)_{R,N_t}} \right]  + \frac{1}{2N_t} \tr\left[\ln{(-\Delta )_{R,N_t}} \right],
\end{equation}
which is the desired result.

\bibliography{u1string}

\end{document}